\documentclass[acmlarge]{acmart}

\usepackage{multicol}
\usepackage{makecell,multirow}
\usepackage{subfig}

\AtBeginDocument{%
  \providecommand\BibTeX{{%
    \normalfont B\kern-0.5em{\scshape i\kern-0.25em b}\kern-0.8em\TeX}}}

\setcopyright{acmcopyright}
\copyrightyear{2018}
\acmYear{2018}
\acmDOI{10.1145/1122445.1122456}

\acmJournal{POMACS}
\acmVolume{37}
\acmNumber{4}
\acmArticle{111}
\acmMonth{8}

\begin{document}

\title{Novel Computational Linguistic Measures, Dialogue System and the Development of SOPHIE: Standardized Online Patient for Healthcare Interaction Education}
\author{Mohammad Rafayet Ali,
        Taylan Sen,
        Benjamin Kane,
        Shagun Bose,
        Thomas M Carroll, 
        Ronald Epstein,
        Lenhart Schubert,
        Ehsan Hoque}

\renewcommand{\shortauthors}{Ali, et al.}

\begin{abstract}

In this paper, we describe the iterative participatory design of SOPHIE, an online virtual patient for feedback-based practice of sensitive patient-physician conversations, and discuss an initial qualitative evaluation of the system by professional end users. The design of SOPHIE was motivated from a computational linguistic analysis of the transcripts of 383 patient-physician conversations from an essential office visit of late stage cancer patients with their oncologists. We developed methods for the automatic detection of two behavioral paradigms, lecturing and positive language usage patterns (sentiment trajectory of conversation), that are shown to be significantly associated with patient prognosis understanding. 
These automated metrics associated with effective communication were incorporated into SOPHIE, and a pilot user study identified that SOPHIE was favorably reviewed by a user group of practicing physicians.
\end{abstract}



\maketitle
\section{Introduction}
Effective patient-physician communication is fundamental to a patient's right to be fully informed and actively involved in health decision making. Good communication skill facilitates physicians' understanding of patients' symptoms, concerns, and treatment wishes \cite{back2005approaching}. Effective communication in the clinical setting has further been correlated with better patient health outcomes \cite{kaplan1989assessing,nawar2007national,oates2000impact,beck2002physician}. Alternatively, a lack of effective communication often results in patients underestimating their disease severity \cite{weeks2012patients} and overestimating their prognosis \cite{gramling2016determinants}. As a result, training the physicians on the fundamentals of how to communicate with patients, including taking turns, asking questions, showing empathy, and being positive is very important part of medical education. The state of the art medical education involves using trained actors who play the role of standardized patients and provide feedback, as well as in person training. These techniques have the limitations of being expensive in terms of time and money as well as being prone to individual variation. Medical schools at the developing world may not even have the resources to provide such training \cite{stoltenberg2020central,van2020increased}. There exists a dire need to improve patient-physician communication training that is not only evidence-based and standardized but also is rapidly customizable, cost-effective, and ready to be deployed online across geographical boundaries.      

The 2020 pandemic saw a dramatic increase in the ubiquity of online interaction. In-person communications were aggressively replaced with virtual interactions across a wide spectrum of domains from education to healthcare. Even the most technologically inexperienced and averse, from preschool students to senior citizens, were forced to learn and find the means to participate online. The medical education system also felt pressure to accelerate physician training, as medical students in Europe and United States were graduated early in order to join health care workers on the front lines \cite{rush}. Ominously, this rush in medical training, together with the loss of in-person interaction, may be likely to exacerbate the current deficiencies in patient-physician communication. This problem is further complicated by the ever decreasing amount of time physicians have to spend with their patients. Additionally, with increasing medical technologies to learn and ever more specialized fields of medical training, physicians have less and less time for training in patient-physician communication. 

In this paper, we focus on 'cancer care' as the conversation context. Communication between oncologists and patients is especially important due to the complexity and the emotion involved in discussing the patients life expectancy. In addition, the oncologists need to explain the severity of cancer, the multiple treatment options available, and the correlates of patient involvement in complex decision-making \cite{millard2006nurse,vahdat2014patient}. Identifying modifiable correlates of effective communication could lead to improvements in physician communication training that engender improvements in patient outcomes. Despite decades of communication training and research, studies have shown that over 60\% of late stage cancer patients come out of their appointments not understanding their prognosis \cite{epstein2017effect}. 

\begin{figure}
    \centering
    \includegraphics[width=12cm]{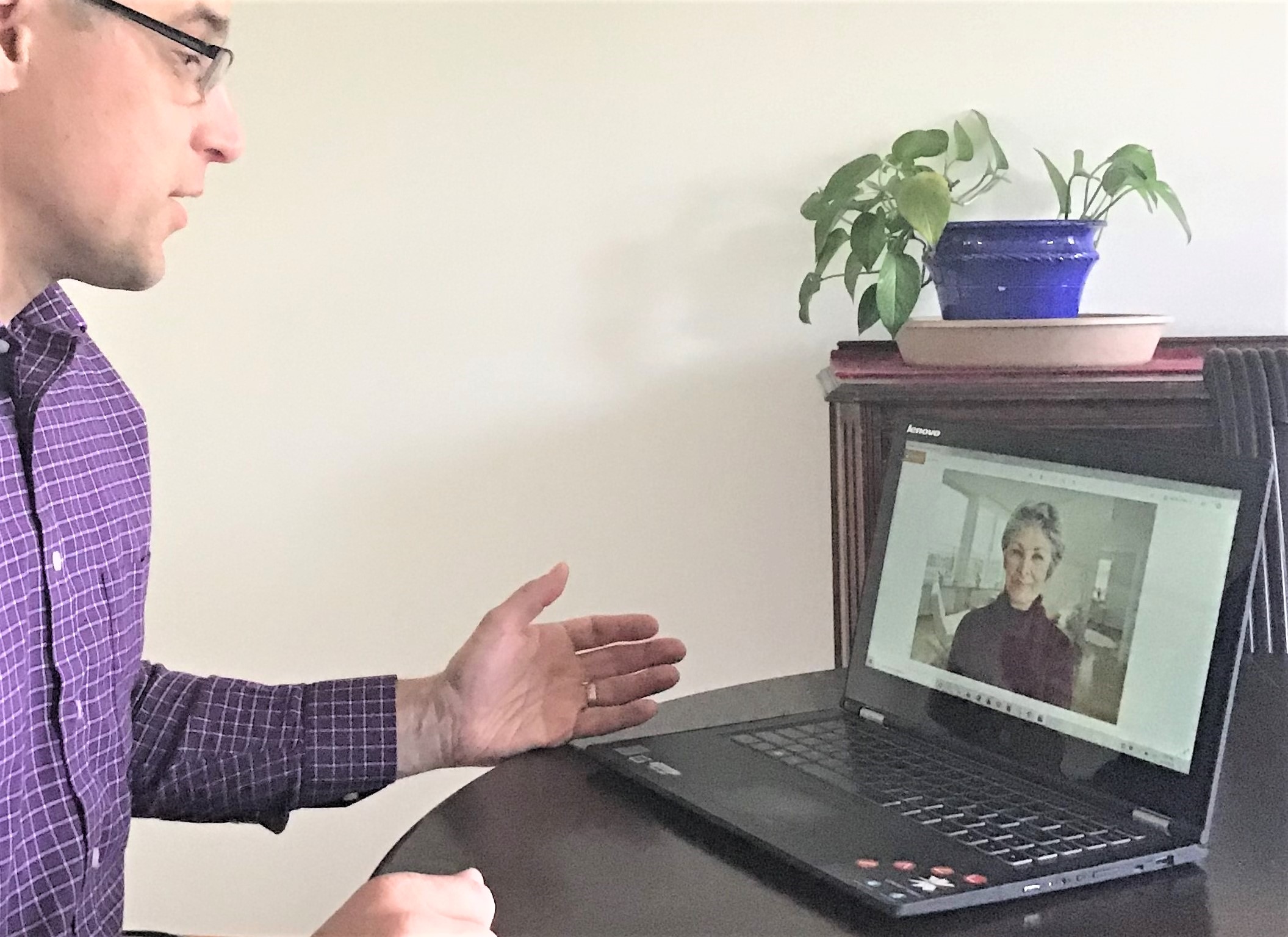}
    \caption{A physician practicing communication skills with SOPHIE virtual patient.}
    \label{fig:participant}
\end{figure}

In this paper, we present a multi-stage research project leading to the development and pilot study of an online virtual patient for training physicians to be better communicators. We begin with the development of automatic detection methods of two behavioral paradigms, \textit{lecturing} and positive language usage patterns (the \textit{sentiment trajectory} of conversation), that are important for patient-physician communication. We have used a data set that consists of 382 transcripts of conversations between late stage cancer patients (Male=172, Female=210) and their physicians (Male=25, Female=13) and a measure of each patient's prognosis understanding \cite{gramling2016determinants}. All conversations involved a regularly scheduled essential office visit. All patients were late-stage (stage 3 or 4) cancer patients. Computational linguistic analysis of the conversation transcripts enabled us to develop automatic metrics for evaluating the degree of lecturing-like structure a conversation has. In addition, we identify that most physicians tend to use one of three styles of varying their linguistic tone over time (i.e. there are three styles of sentiment trajectory). Further, we show that these metrics have a significant association with patients' level of prognosis understanding. We then developed an online virtual agent-based communication skills development system, SOPHIE, which gives users feedback on lecturing and positive language usage. SOPHIE also provides feedback on the user's speech rate and number of questions asked. SOPHIE presents herself as a late stage cancer patient. SOPHIE was designed following the well established physician communication training protocol -- SPIKES \cite{baile2000spikes}, to enable the physicians practice their communication skills focusing on prognosis understanding. Fig. \ref{fig:participant} shows a physician practicing communication skills with SOPHIE in his home. The ubiquitous nature of online technologies, such as SOPHIE, allows users to practice in their own private environment. In this paper, in collaboration with oncologists and medical educators from a leading medical institution of North America, we provide early ideas on how inspirations from UbiComp could potentially transform current medical education.     


Our contributions include:
\begin{itemize}
    \item The development of an automated metric for measuring the lecturing-like structure, and the identification that most doctors use one of three styles of sentiment trajectory (i.e., pattern of modifying their positive language usage over the course of a patient-physician conversation).
    \item Demonstration that the degree of lecturing structure is significantly associated with the level of prognosis misunderstanding. 
    \item Finding that a certain sentiment trajectory style (one which involves delivering technical information and ending with positive language) is associated with better prognosis understanding. 
    \item Presentation of an iterative participatory design process, and an initial end user evaluation with eight practicing physicians, of an online virtual patient (SOPHIE - Standardized Online Patient for Healthcare Interaction Education) for feedback-based practice of critical patient-physician conversations. 
\end{itemize}

\section{Related Work}
This work touches on several interconnected areas including, affect and sentiment analysis, prognosis understanding, patient-physician communication, virtual patients, and communication skills development programs. Here we touch on the related research on above mentioned topics. 

The association of positive patient outcomes with physician affect has received limited experimental examination. Within the limited studies that have been conducted, differing conclusions have been reached in regard to the association of physician sentiment with patient health outcomes. In a review of 22 studies involving patient-physician communication, Beck, et al. \cite{beck2002physician} found that expressions of "good feelings" by the physician "in regard to patient’s actions" have a direct association with patient health outcomes. Similarly, Di Blasi, et al. \cite{di2001influence} found in their systematic review of 25 randomized controlled trials on affective physician communication, that even though studies show inconsistency regarding emotional and cognitive care. However, authors found that physicians who adopt a warm, friendly, and reassuring manner are more effective than those physicians who keep consultations formal and do not offer reassurance. In contrast, Kaplan et al. \cite{kaplan1989assessing} found that positive affect expressed by physicians during baseline visits was not associated with better patient health status at a follow-up visit. Sen et al. \cite{sen2017modeling} also found a lack of association of overall conversational positive sentiment with patient ratings of their oncologists communication skill.

Prior studies have also studied the association of physician affect on patient information recall and prognosis understanding. In a study of women receiving simulated breast cancer-related communications from a videotaped oncologist, van Osch et al. \cite{van2014reducing} found that affective communication improves information recall. When physicians used positive affect statements (such as "We will do and will continue to do our very best for you.", "Whatever happens, we will never abandon you. You are not facing this on your own.") participants provided significantly more correct answers on a questionnaire testing the participants’ recall of details in the diagnosis, prognosis, and treatment options. A similar study involving participant viewing of videotaped simulated oncologist communications, Shapiro et al. \cite{shapiro1992effect} found that participants who received communication from a worried physician as opposed to the standard, recalled significantly less medical information. Another related study by Fogarty et al. \cite{fogarty1999can} found that women viewing videotaped interactions of physicians with "enhanced compassion", as compared to standard physicians, recalled significantly less information.

What could be causing these inconsistent results? The importance of sentiment variation over time has long been recognized in story telling \cite{vonnegut1999palm} and more recently has been shown to be relevant in natural language analysis \cite{trilla2012sentence}. Ali et al. \cite{ali2018and} demonstrated the importance of timing of emotional expressions in dyadic conversation. Reagen et al. \cite{reagan2016emotional} applied natural language processing techniques to analyze 1327 written stories and identified that there exist six common emotional trajectory styles. An analysis of textual sentiment trajectory was applied to 27,333 YouTube vlogger (i.e. an individual who actively provides video logs on various topics) videos by Kleinberg et al. \cite{Kleinberg}. They identified that seven common trajectory styles exist, and found videos with the most view count have a style which ends with a high positive sentiment. 

In health care communication skills training, virtual agents and online platforms have been used in an attempt to provide an effective and reproducible experience. Liu et al. (\cite{liu2016eqclinic}) developed the EQClinic platform for medical students. Their tool provided summary reports about speaking contribution, volume, and pitch as well as facial expressions, head positioning/nodding, and hand-over-face. In a study with medical students, authors found that reviewing summaries of non-verbal communication behaviors collected by EQClinic improved students' interview skills. DeVault et al. (\cite{devault2014simsensei}) developed SimSensei - a virtual agent in the context of a healthcare decision support system. The goal of this system is to identify psychological distress indicators through a conversation with a patient in which the patient feels comfortable sharing information. This system has both nonverbal sensing and a dialogue manager. The dialogue manager uses four classifiers to categorize the user's speech, and hence to generate a relevant response. Peddle et al. (\cite{peddle2019development}) developed a virtual patient (VP) in a virtual reality platform. The aim of the program was to develop the nontechnical knowledge, skills, and attitudes among undergraduate health professionals, and to enable them to practice the non-technical skills. The VPs in the program belong to the narrative category portraying a patient's story as it evolves over time, with a cause-and-effect approach to demonstrate consequences of actions and decisions. In a study with second and third year nursing students, the authors found that interactions with VPs developed knowledge and skills across all categories of non-technical skills to varying degrees. Third-year students suggested that interactions with VPs helped develop knowledge and skills in a clinical setting. Web-SP (\cite{zary2006development}), is a web-based platform, designed to facilitate the development of virtual standardized patients (SP) for healthcare education. It provides a general platform to design and develop virtual patients for specific healthcare cases. Angus et al. (\cite{angus2012visualising}) developed a graphical visualization tool to model patient-physician dialogue to identify patterns of engagement between individuals including communication accommodation, engagement, and repetition. Kleinsmith et al. (\cite{kleinsmith2015understanding}) developed a chat-based interactive virtual patient for early stage medical students to practice empathetic conversation. During the training, students can gather information regarding the history of the present illness, medical history, family history and social history. Additionally, during each session, the VPs delivered a statement of concern. These statements, termed empathetic opportunities, were designed to elicit an empathetic response from the user. In a study, medical students interacted with the VP and standardized patients. The responses of the participants were then rated by coders, and it turned out that responses were more empathetic with virtual patients than with standardized patients. The level of empathetic response positively correlated with response length. Bond et al. (\cite{bond2019virtual}) used virtual agents to train and generate cases for a history-taking task among resident physicians. The system gives a score to the physicians after performing the history taking. 

In this work, we have focused on improving prognosis understanding among late-stage cancer patients. To this end we designed a virtual patient to conduct conversations with oncologists. To provide feedback to users on communication skills, we first developed algorithms to detect behavioral cues, and then engaged practicing physicians in a participatory design to refine the program's feedback module.    

\section{Materials}
We performed a post-hoc analysis of a study (\cite{hoerger2013values}) involving 382 visits between cancer patients ($N=382$) and their oncologists ($N=38$). The data includes a transcript of the conversation, in addition to both patient and physician surveys associated with each visit. The survey also included questions to the physician and to the patient regarding the patient's prognosis. The prognosis questions were a modified version of the SUPPORT prognosis measure (\cite{weeks1998relationship}). Specifically, the prognosis question directed to the physicians was: \textit{"What do you believe are the chances that this patient will live for 2 years or more?"}; the options provided for a response are shown in Table \ref{table_UPresponses}. 
\begin{table}[t]
\caption{Study Data: Counts and Prognosis Survey Options}

\centering
\begin{tabular}{c|l}
\label{table_UPresponses}

Resp. \# & Description  \\
\hline
0	& 100\%       \\   
1	& about 90\%  \\
2	& about 75\%  \\
3	& about 50-50 \\
4	& about 25\%  \\
5	& about 10\%  \\
6	& 0\%         \\ 
X	& don't know  \\
X	& refuse to answer     \\
\hline
\end{tabular}

\end{table}

Patients were separately asked \textit{"What do you believe your doctor thinks are the chances that you will live for 2 years or more?"}, with the same options for a response. By comparing patient and physician responses, we derived a misunderstanding percentage. More specifically, when the absolute difference of the responses is greater than 1, the patient-physician prognostic understanding is defined as being misunderstood. Data in which either the physician or patient refused to answer were not used. The transcribed visits each involved a regularly scheduled visit between a late-stage (stage 3 or 4) cancer patient and their oncologist. Many of the visits included a family caregiver and/or other health care staff (e.g., nurse, second physician).

\section{Methods}

In patient-physician communication there are several behavioral paradigms that help improve information transfer. Our focus is on automating the identification of behaviors that help to improve prognosis understanding. Our goal is to use those behaviors to provide feedback to the oncologists when they practice conversations with standardized or virtual patients. The two categories of behaviors of interest in patient-physician communication are behaviors to avoid, and behaviors to cultivate. Among many behavioral paradigms, we have explored two patterns of behavior -- \textit{lecturing}, and the \textit{sentiment trajectory} of conversation. We first present how we set about detecting these phenomena automatically and determining how they are associated with prognosis understanding. Then we explain our feedback design for these two behavioral patterns, applicable  in conversation practice with a virtual conversational agent.  

\begin{figure}
\centering
\includegraphics[width=\textwidth]{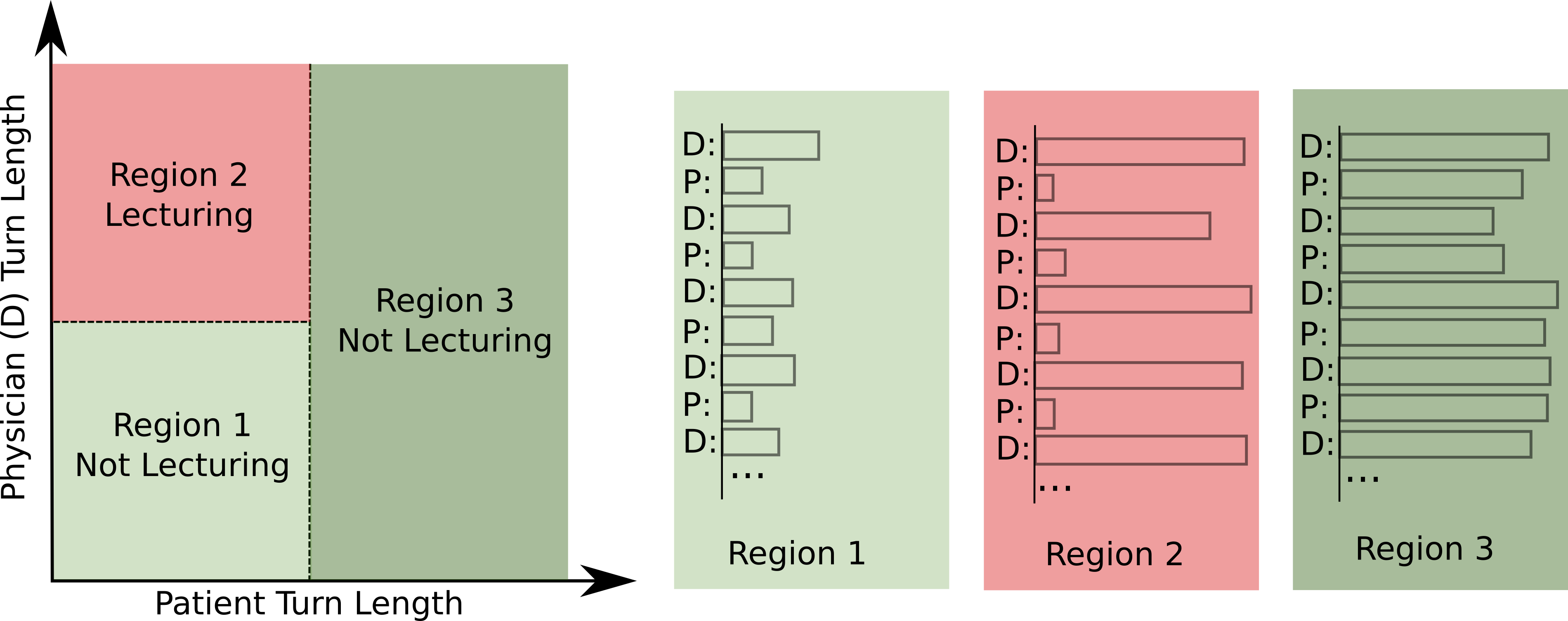}
\caption{Classification of Physician (D) and Patient (P) Turn Lengths over Window as Lecturing and Not-Lecturing a) Regions of Lecturing and Not-Lecturing as P length vs. D length b) example transcript turn lengths}
\label{fig:LecturingAlgorithm}
\end{figure}

\textbf{Lecturing:} Lecturing occurs when the physician delivers a lot of information without giving the patient a chance to ask questions or to respond (\cite{siminoff2000doctor}). In order to detect a lecturing event, we calculate the turn lengths (i.e., number of words spoken by the physician or the patient) in a fixed-size sliding window. When the number of words spoken by the physician is greater than a threshold, we classify it as a lecturing event. Fig. \ref{fig:LecturingAlgorithm} shows the area where the lecturing event can occur. The thresholds are determined by maximizing the entropy of the outcome variables (i.e., prognosis misunderstanding). In this section, we present a method of detecting lecturing in dialogues. Later we present the relationship between the lecturing and prognosis misunderstanding.  

\textbf{Sentiment Trajectory:} 

In the field of natural language processing, \textit{sentiment} represents the classification of emotion in text data. In this work, we focus on positive language usage. We define sentiment trajectory as the change that occurs in physician positive sentiment over the course of the conversation. 
Findings from communication research suggest that the trajectory of affective communication features (e.g., sentiment) may be particularly important \cite{ali2018and}. However, the physician sentiment trajectory over a conversation has not been well-studied in the context of patients’ prognosis understanding. First we describe how we have defined and identified effective sentiment trajectories. Later we present the association between the trajectories and prognosis understanding.
\subsection{Lecturing}
\label{sec:lecturing}
Here we describe our automated algorithm for calculating the LECT-UR Score (Lecturing Estimation through Counting Turns with an Unbalanced-length Ratio), a measure of lecturing-related conversational structure. The LECT-UR Score is based on Back, et al. \cite{back2005approaching}'s definition of lecturing (i.e., when a Patient-physician transcript shows turns when the "physician delivers large chunks of information without giving the patient a chance to respond or ask questions") \cite{back2005approaching, siminoff2000doctor}. The LECT-UR scoring technique was not "trained" on a set of subjective, manually labeled instances of human perceived "lecturing", but rather was designed as an objective algorithm to measure lecturing-related conversational structure, i.e., how often the physician turns were disproportionately longer than his/her patient's turns over a window of conversational turns.


As shown in Fig. \ref{fig:LecturingAlgorithm} region 1, when both the physician and patient speak with short duration over a window of conversation, it is not counted as an instance of lecturing. Similarly, if, as shown in region 3, the patient is speaking with a long average turn length over the window, it is not labeled as lecturing. Only when the physician's average turn length exceeds a threshold level over the patient's average turn length, shown in Region 2, is the window labeled as an instance of lecturing.

This algorithm is expressed in the following equations:
\begin{equation}\label{eqn:lect_algo}
\mathbb{L} = \sum_{\forall_k} \mathbb{I}\left(\sum_{i=k w_i \in D}^{k+\mathbb{W}}\omega_i - \tau \right) \times \mathbb{I}\left(\tau - \sum_{i=k w_i \in P}^{k+\mathbb{W}}\omega_i  \right)  
\end{equation}

\[   
\mathbb{I}(x) = 
     \begin{cases}
       0 :  x < 0\\
       1 : x \geq 0\\ 
     \end{cases}
\]

where,

\quad $\mathbb{L}$ : LECT-UR Score

\quad $\mathbb{W}$ : window length in number of turns

\quad $\tau$ : turn length disparity threshold

\quad $\omega$ : words in the transcript

\quad D : physician utterances

\quad P : Patient utterances
\\

Referring to equation \ref{eqn:lect_algo}, a value for the $\tau$ parameter must be determined. As $\tau$ approaches zero, the area of region 1 in Fig. \ref{fig:LecturingAlgorithm} will also approach zero. Alternatively, if a very large value is used for $\tau$, every window will be classified as \textit{not lecturing} since region 1 will cover the entire data space. In order to be useful, the LECT-UR score should have variability (i.e. if all data points have the same LECT-UR score, we will not learn much). Borrowing concepts from information theory, the amount of \textit{information} in a signal can be measured by the signal's \textit{entropy}, where entropy is a measure of the amount of disorder or uncertainty \cite{shannon2001mathematical}. More specifically, for a given data set $X$, the definition of the entropy, $H(X)$, is:

\begin{equation}\label{eqn:entropy}
\\
H(X) =  \sum_{i=1}^n {\mathrm{P}(x_i) \log_b \frac{1}{\mathrm{P}(x_i)}}
\\
\end{equation}

where $P(x_i)$ represents the probability of observing the $i^th$ data point. As the probability of an event $x_i$ approaches certainty (i.e. $P(x_i) \approx 1$), the information content approaches zero. Similarly, as the probability of an event $x_i$ approaches zero, the contribution of such events to the total information content in the data set approaches zero. Thus, in order to maximize the information contained in the LECT-UR score, the scores should be well distributed (i.e. maximizing the entropy). 

\begin{figure}[tb]
\centering
\subfloat[]{\includegraphics[width=7cm]{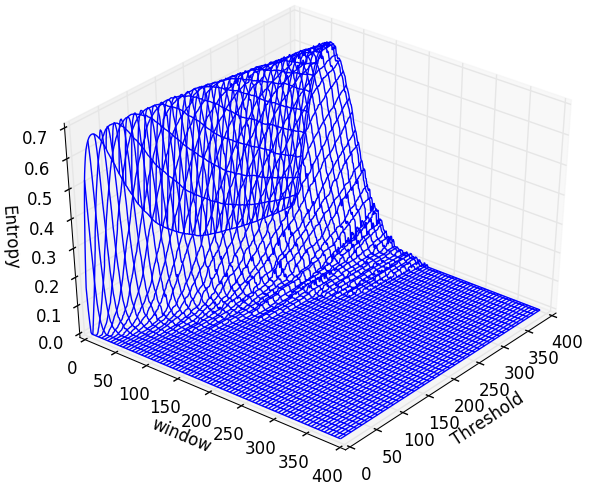} \label{fig:EntropyvsWindowvsTauContour}}
\subfloat[]{\includegraphics[width=7cm]{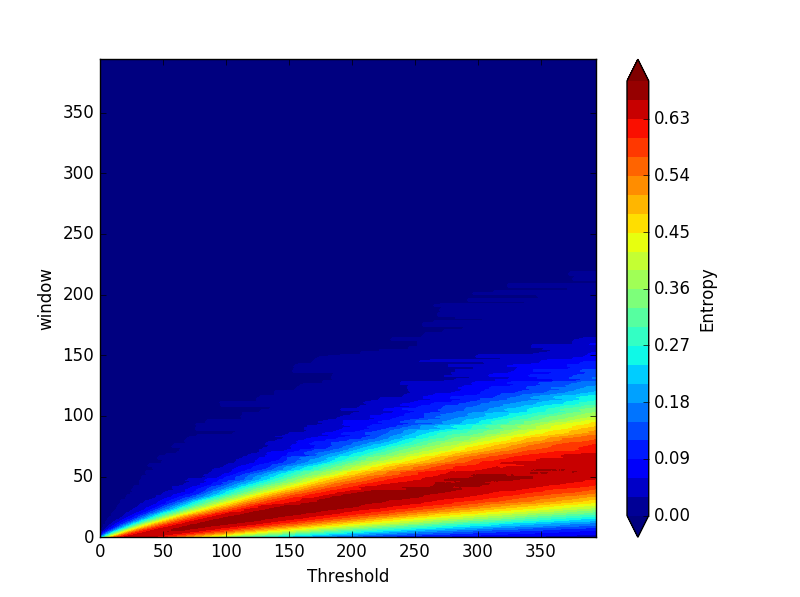}
\label{fig:EntropyvsWindowvsTau}}
\caption{Finding the Optimal Lecturing Threshold and window size Based on Entropy. a) Contour plot of Entropy, b) Heatmap of Entropy}

\end{figure}

In order to determine the optimal $\tau$ and $\mathbb{W}$, we perform a grid search. For a given $\tau$ and $\mathbb{W}$ we first calculate the LECT-UR score $\mathbb{L}$ based on equation \ref{eqn:lect_algo}. We then applied the kernel density estimation method \cite{parzen1962estimation} to compute the probability density function $P(x)$. From the probability density function we then obtain the entropy of $\mathbb{L}$ using equation \ref{eqn:entropy}. In Figs. \ref{fig:EntropyvsWindowvsTau} and \ref{fig:EntropyvsWindowvsTauContour}, the entropy values for different values of $\tau$ and $\mathbb{W}$ are shown. The maximal entropy occurs with $\tau = 103$ and $\mathbb{W} =20$. After calculating the LECT-UR score with the optimal parameters for each office visit transcript, we partition the data into high and low LECT-UR groups based on the median value. We then use the Z-score two-tailed population proportion test to see the difference in the percentage of prognosis misunderstanding.

\subsection{Sentiment Trajectory}
To investigate the relevance of speaking with positive sentiment as part of an automated system, we utilized the VADER (Valence Aware Dictionary for sEntiment Reasoning) automatic text analysis tool (\cite{hutto2014vader}). VADER calculates sentiment through the use of a rule-based model that employs a sentiment lexicon (dictionary of words containing an associated valence measure). The sentiment lexicon used by VADER was produced from a human-labeled corpus in which humans rated sentiment in terms of the overall positive, neutral, or negative emotion associated with a given word in a phrase or sentence. The VADER positive sentiment feature is the result of a large number of human raters' understanding of positive and negative emotion associated with particular words. The VADER positive sentiment score was evaluated for each turn of the conversation. These physician and patient sentiment scores were used in two ways —- 1) average analysis, and 2) sentiment trajectory.

In average sentiment analysis, the average sentiment scores for the physician were calculated for each transcript. The transcripts were split into two groups based on the median of the physician average sentiments (i.e. a High Sentiment group and a Low Sentiment group). The outcome measure (Prognosis Misunderstanding\%) is then compared between the two groups using the z-score population proportion test.

For the second way of using physician and patient sentiment scores, we defined the sentiment trajectory as the time series of average physician positive sentiment over the segmented conversation. More specifically, we partitioned each conversation transcript into a number of non-overlapping segments (each segment having the same number of conversational turns) and calculated the physician's average positive sentiment within each segment. Each conversation's sentiment trajectory is represented as a multidimensional vector, each dimension corresponding to the average sentiment within a corresponding segment of the conversation.

We next determined whether distinct styles of physician sentiment trajectory existed among the conversations and investigated whether any of these physician styles demonstrated significant differences in any of the indicators of communication effectiveness. To determine whether distinct styles of sentiment trajectory exist in the physician sentiment among the transcripts, we applied the k-means clustering algorithm (\cite{lloyd1982least}). The k-means algorithm groups the conversation trajectories into a number (k) of clusters (or groups) of trajectories based on their relative Euclidean distance. The number of clusters k was selected using the widely used Silhouette method (\cite{rousseeuw1987silhouettes}), in which a grid search over a finite space of integer values for the k parameter is performed in order to find the number of clusters which maximizes the Silhouette score (i.e. a combined measure of cohesion among data points within a given cluster and separation of data points among different clusters). In order to determine whether any of the resulting sentiment trajectory clusters had statistically significant differences in the outcome measures, we applied the inference test for population proportions pairwise between the groups. 

To understand the effects of the demographic and confounding variables we performed a logistic regression analysis. Specifically, we applied logistic regression on gender, age, disease severity, average sentiment of the conversation, study site, study arm, and the conversation styles to predict the outcome measures. In analyzing confounding variables, there are mainly two approaches: 1) stratification and 2) multivariate methods (i.e., logistic regression aka logit). Since we have multiple potentially confounding variables, we used the multivariate method of logistic regression. Stratification would likely be problematic since we have a small sample size. Our outcome variable is binary (either you understand or don't understand your prognosis), therefore instead of linear regression we use logistic regression, which estimates the probability of getting an outcome as a linear function of all of the input variables (including confounding variables as well as cluster membership). 


After fitting data to logistic regression, we can compare the relative effect that each of the input variables has on predicting whether a given data point (conversation) results in a "Don’t understand prognosis" classification. After normalizing the inputs (i.e., scaling and shifting to have mean=0 and variance=1) we fit the model (using the hyper-parameter that provides the highest data likelihood) and hence find the model weights. We then investigate the weights of the logistic models and the prognosis misunderstanding percentage for each of the conversation styles.  
\section{Findings}
\subsection{Association between LECT-UR Score and Prognosis Understanding}
\begin{figure}
\centering
\includegraphics[width=11cm]{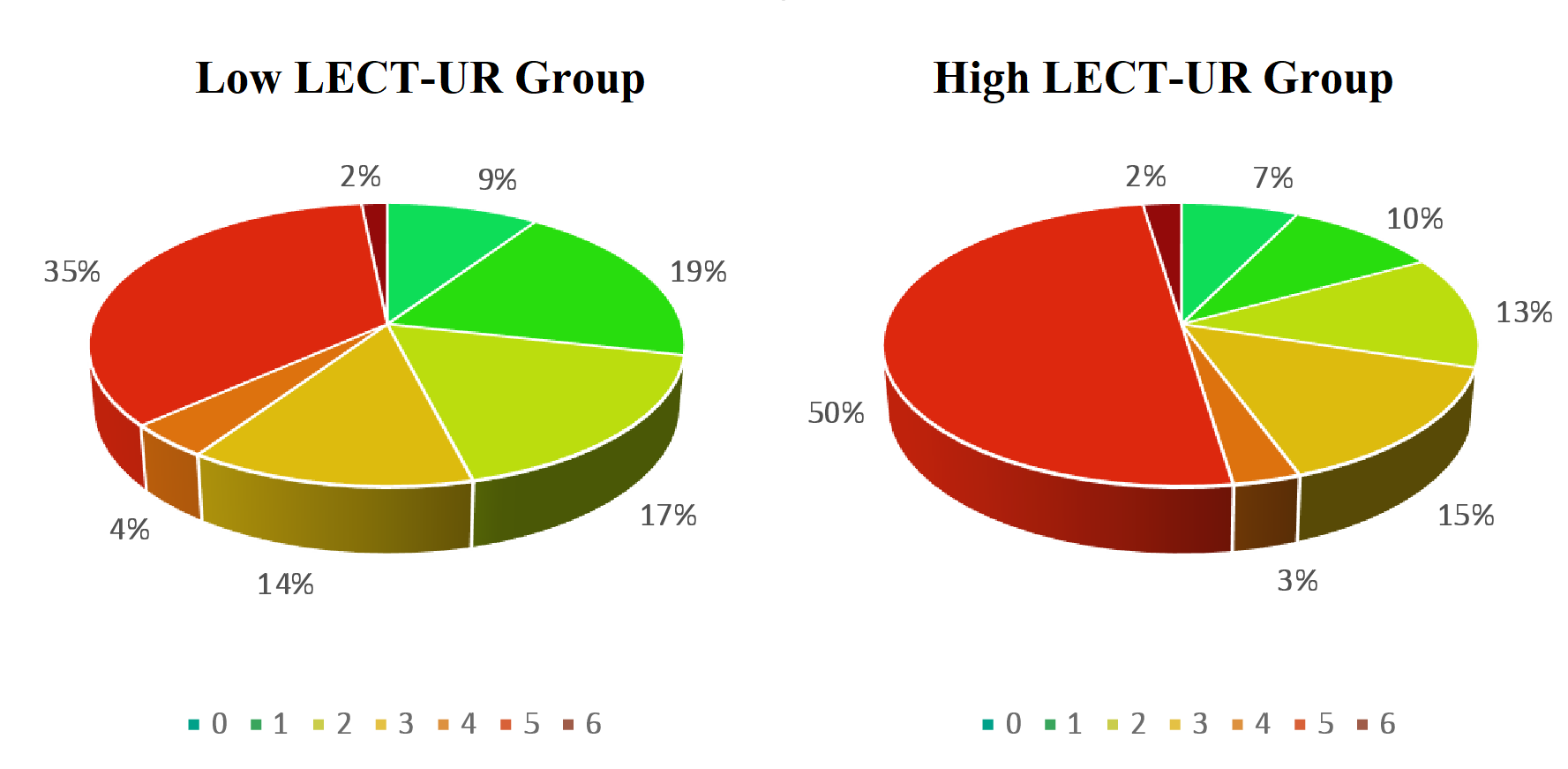}
\caption{Distribution of Prognosis Misunderstanding Levels in the High and Low LECT-UR Groups}
\label{fig:DiscordanceDistributionInLecturing}
\end{figure}

As shown in Table \ref{TableDiscordance}, the High LECT-UR Score group has a larger percentage of prognosis misunderstanding than the Low LECT-UR Score group (83.6 vs. 72.3) with a corresponding p-value of 0.00058 and an estimated Cliff's d effect size of 0.37 \cite{cliff1993dominance}. 
Shown in Fig. \ref{fig:DiscordanceDistributionInLecturing} are the distributions of the Prognosis Misunderstanding for the high and low LECT-UR groups. Note that we have calculated the misunderstanding percentage based on the absolute difference between the survey question response. Here a difference of 5 and 6 were treated as misunderstood prognosis. In the high LECT-UR group, over 50\% of the patients had a prognosis misunderstanding of 5 or 6 \textit{levels}. A level of 5 or 6 represent the situations in which there is a 90\% or more difference between the patient's understanding of their physician's two year survival estimate and their physician's actual two-year survival estimate.

\begin{table}
\centering
\caption{Average Prognosis Misunderstanding scores in High and Low LECT-UR, Average Sentiment, and Sentiment Synchronicity Groups}
\begin{tabular}{lcrrr}
\label{TableDiscordance}
Group & \makecell{Prognosis \\ Misunderstanding \%}  & p-value & effect size \\
\hline
\makecell[l]{High LECT-UR\\Low LECT-UR} & \makecell{83.6\\ 72.3}  & 0.00058  & 0.37\\
\hline

\end{tabular}
\end{table}

\subsection{Association between Sentiment and Prognosis Understanding}

\begin{figure}
    \centering
    \includegraphics[width=8cm]{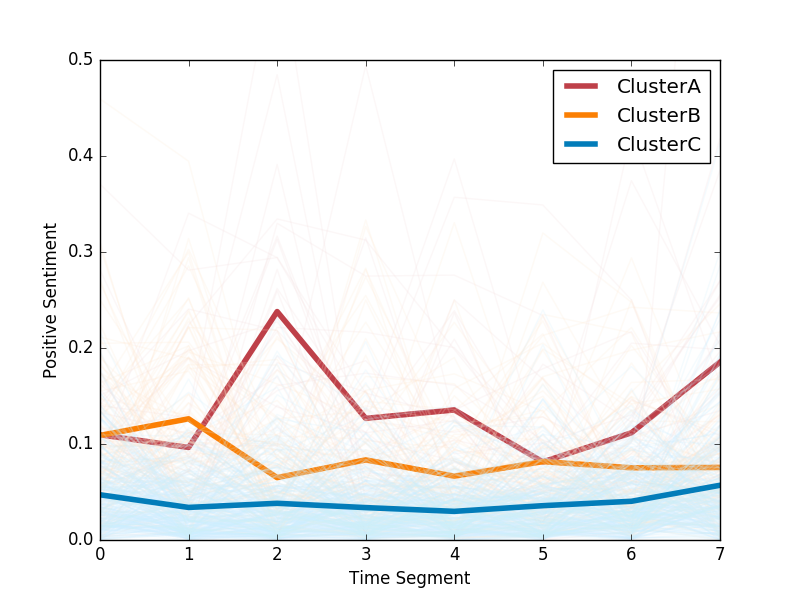}
    \caption{Resulting Sentiment Trajectory Clusters for best K=3. }
    \label{fig:clusters}
\end{figure}

The difference in the prognosis misunderstanding \% between the high and low average positive sentiment groups did not show a significant difference. Out of the analyzed number of clusters (k = [2, 10]), the number of trajectory clusters that had the highest silhouette score was k=3. In addition, the BIC (Bayesian information criterion \cite{schwarz1978estimating}) analysis also found that the optimal value for k is 3. Shown in Fig. \ref{fig:clusters} are the resulting three trajectory clusters: cluster A (red, n = 15); cluster B (orange, n = 58), and cluster C (blue, n = 191). It should be noted that the K-means clustering algorithm does not inherently attempt to produce clusters of equal sizes, but rather finds clustering groupings which minimize the within-group variation. Cluster A (Dynamic) is represented by a more dynamic shape with increases in positive sentiment at 25\% into the conversation (segment 2), as well as at the end of the conversation (segment 7). In contrast, Clusters B (Medium) and C (low) have a mostly flat positive sentiment level throughout the conversation with approximate average VADER sentiment levels of 0.1 and 0.05 respectively. 

Shown in Table \ref{tab:PMU_sentiment} are the outcome measures for each of the three trajectory cluster groups along with pair-wise population percentage inference test p-values.  As shown, the Prognosis Misunderstanding \%, the low cluster (cluster C) showed the highest percentage with 67.9 \% of the patients having a discordant understanding of their prognosis. The p-values for comparing the percentages between low and dynamic and low with medium clusters were 0.04 and 0.06 respectively.

\begin{table}[]
    \centering
    \caption{Prognosis misunderstanding \% in three sentiment trajectory clusters.}
    \begin{tabular}{c|c|c|c|c|c}
    \hline
         \multicolumn{3}{c|}{Trajectory Group (size)} & \multicolumn{3}{c}{Pairwise statistical comparison}\\\hline
           A (15)& B (58) & C (191)& $P_{AB}$ & $P_{BC}$ & $P_{AC}$ \\\hline
          	46.1 &	52.6 &	67.9 & 0.34 & 0.04& 0.06\\\hline
    \end{tabular}
    
    \label{tab:PMU_sentiment}
\end{table}

\begin{figure}
    \centering
    \includegraphics[width=8cm]{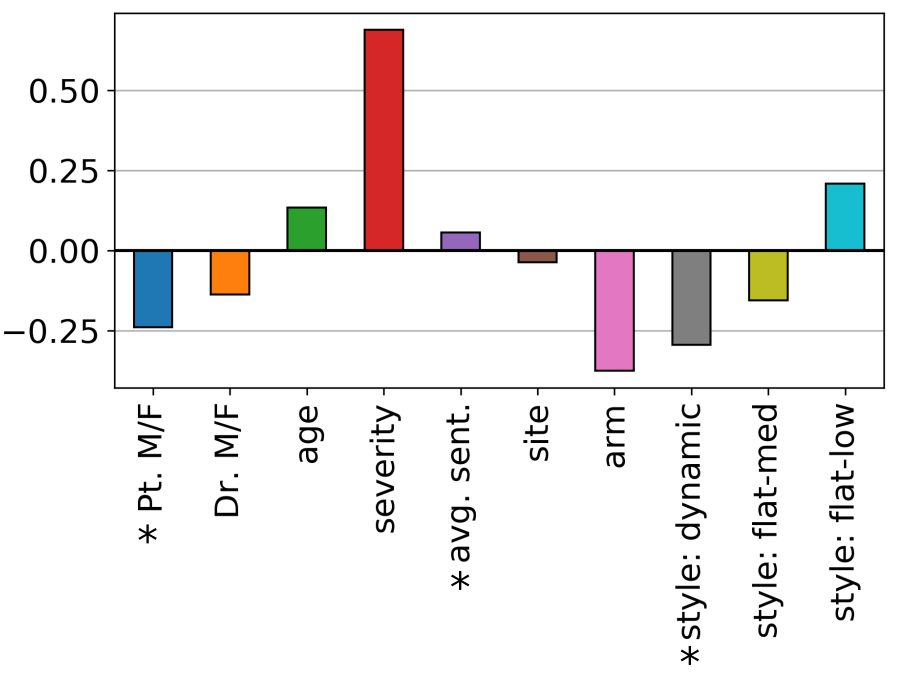}
    \caption{Logit model weights for predicting whether the prognosis misunderstood.}
    \label{fig:logit_weights}
\end{figure}

Fig. \ref{fig:logit_weights} shows the logistic regression weights when predicting the Prognosis Misunderstanding \%.  The variables marked with a (*) had p<0.05. The more positive weights indicates higher chances of the particular outcome. In fig. \ref{fig:logit_weights} the higher positive value was assigned to severity and age by the trained model. In Fig. \ref{fig:logit_weights} the highest positive value was assigned to severity. This indicates that patients with higher severity level of disease are more likely to misunderstand their prognosis. Among all the clusters the dynamic cluster has the lowers (negative) value. This indicates that when physicians used the dynamic sentiment throughout the conversation the patients were less likely to misunderstand their prognosis.  It should be noted that in both outcome measures the effect of average positive sentiment variable is very small. 

Unlike the linear regression, with logistic regression there is not a simple way to adjust the output (i.e. “correct” the output) for the effect of confounding variables of each data point. This is because the actual outputs are binary, whereas the model output is a probability. Instead, we can compare the predicted model Prognosis Misunderstanding \% for each cluster. When all confounding variables are set to have the average value over our data set, we compute the models’ predicted Prognosis Misunderstanding \% for each cluster (see table \ref{tab:logit_model}).

The Wald test p-value of the logistic regression is also shown in table \ref{tab:logit_model} (marked * in Fig. \ref{fig:logit_weights}  when p < 0.05). This again indicates that with confounding adjustment the dynamic style cluster has Prognosis Misunderstanding \% among other clusters.

\begin{table}[thb]
    \centering
    \caption{Confounder-Adjusted Logit Model}
    \begin{tabular}{c|c|c|c}
         Group	& PMU \% &	$\beta$ &	p-val\\\hline
         A	 	& 49.76	& -0.294 &	 0.033\\
         B	 	& 70.74	& -0.155 &	 - \\
         C	 	& 84.85	&  0.209 &	 - \\\hline
    \end{tabular}
    
    \label{tab:logit_model}
\end{table}

\section{Design of SOPHIE}
Our aim is to develop a virtual standardized patient for practicing communication skills. In medical education, students practice with a standardized patient -- an actor/actress pretends to have a medical condition. Students interact with the standardized patients and are later they receive feedback on their interaction. Our goal is to allow the medical students to practice their communication skills with a virtual agent, allowing multiple repetitions in their own environment, which is difficult for actual standardized patients. 

\subsection{Scenario}
We have developed a prototype of the SOPHIE program, which allows individuals to have a conversation with a virtual agent concerning prognosis and treatment options. SOPHIE presents herself as a late-stage cancer patient. For a pilot study we selected a particular case for the virtual patient, inspired by a case from another study (\cite{shields2019influence,elias2017social}). We have used the SPIKES protocol to guide the conversation \cite{baile2000spikes}. The SPIKES protocol was developed to train physicians deliver bad news. This protocol has shown success in increasing confidence among oncologists in delivering bad news. The SPIKES protocol has six steps – 1) setting up the interview, 2) assessing patients’ perception, 3) obtaining patients’ invitation, 4) giving knowledge and information to the patient, 5) addressing the patient’s emotion with empathetic responses, and 6) strategy and summary. With SOPHIE, at the beginning of the conversation (SPIKES step 1) SOPHIE introduces herself and mentions that she has lung cancer. Then SOPHIE raises the topic of her sleep pattern at night and asks if she needs to change her pain medication, allowing the physician to assess her perception (SPIKES step 2). She states that her current pain medication, \textit{Lortab}, is not working anymore. After discussing the pain medication, SOPHIE turns attention to her test results, giving the physician a chance to obtain her invitation to talk about more difficult topics (SPIKES step 3), before asking more specifically about prognosis if the physician did not already address it, allowing the physician to give information to the patient (SPIKES step 4). SOPHIE then asks about what her options are, allowing the physician to give empathetic responses (SPIKES step 5). Finally, she follows up by discussing whether chemotherapy remains an option, whether she should focus on comfort care, what the side effects of chemotherapy are (if mentioned), and how to break the news to her family, allowing for the physician to give strategy \& summary (SPIKES step 6).


While designing the scenario, we have kept a few things in mind that are important in end-of-life discussion. 
\begin{itemize}
    \item SOPHIE presents herself as already seeing a physician but finding that her medication is no longer working. She knows that she has cancer but is not certain how much time she has left. 
    \item SOPHIE provides an opportunity to the user to discuss her treatment options, but raises the issue of chemotherapy. 
    \item SOPHIE provides an opportunity to discuss her prognosis.
    \item SOPHIE allows for empathetic responses.  
\end{itemize}

This type of discussion promotes understanding of the patient, gathering information from the patient, discussing critical information, and responding with empathy. 

\subsection{Dialogue System}
The SOPHIE program is built on top of Eta, a general purpose dialogue management framework representing a further development of the LISSA system \cite{Razavi2017ManagingCS,razavi2016lissa,razavi2017managing}. Each dialogue agent built withing the Eta framework defines a flexible, modifiable dialogue schema, which specifies a sequence of intended and expected interactions with the user. The body of a dialogue schema consists of a sequence of formal assertions that express either actions intended by the agent, or inputs expected from the user. These events are dynamically instantiated into a dialogue plan over the course of the conversations. As the conversation proceeds, this plan is subject to modification based on the interpretation of each user input in the context of the agent's previous utterance. For instance, if a planned query to the user has already been answered by some part of a user's previous input, the dialogue manager can skip that query. The dialogue manager can also expand steps into subplans by instantiating sub-schemas in the case of more complex interactions.

Both interpretation of the user's replies and generation of the agent's responses are handled using transduction to and from simple context-independent English sentences called \textit{gist-clauses}. The dialogue manager interprets each user's input in the context of SOPHIE's previous question, using this context to select topically relevant pattern transduction hierarchies to use to interpret the user's response. The context of the previous question is useful for resolving anaphora, ellipsis, and other pragmatic phenomena. The rules in the selected hierarchies are then used to derive one or more gist-clauses from the user's input, containing explicit representations of both statements and questions detected in the user's utterance. For example, if SOPHIE asks ``Do you think I should take stronger pain medication?'' and the user answers ``Yes.'', the gist-clause extracted would be ``I think you should take stronger pain medication.'' If the user replies ``Can you tell me more about how you're feeling?'', the gist-clause extracted would be ``Can you tell me more about your pain?'', having interpreted the question as an inquiry about SOPHIE's pain in the particular context of her question.

As mentioned, the gist-clauses are derived using hierarchical pattern transduction methods. Each transduction hierarchy specifies patterns at its nodes that are to be matched to input, with terminal nodes providing result templates to be used according to various directives (e.g. storing as a gist-clause, outputting the result, specifying a sub-schema to be instantiated, etc.). The pattern templates look for particular words or word features, including ``wildcards'' matching any word sequence of some length. In the case of a failure to match, the system first tries siblings of the pattern before backtracking to the previous level, though the efficiency of the hierarchical pattern matching approach lies in the fact that higher levels can segment utterances into meaningful parts, thus reducing the amount of backtracking necessary to interpret the user's input.

The agent's responses to the user are likewise determined using hierarchical pattern transduction. In the case where the gist-clause from the user's utterance is a simple statement, the agent selects a reaction to the gist-clause and either instantiates a sub-schema to ask a follow-up question, or proceeds to the next topic in the main schema. If the gist-clause from the user's utterance is a question, the agent instantiates a sub-schema to select a reply to the user's question and await either a follow-up question or closure from the user. The system also has the potential to form replies to multiple gist clauses from a single user utterance, for instance reacting to the user's statement before responding to a final question by the user.

The transduction hierarchies themselves were designed in a modular fashion, with a ``backbone'' of transduction trees detecting general questions that SOPHIE might expect a user to ask, with additional transduction trees for detecting questions and replies specific to the current topic of the conversation. In the case of a failure to match a specific response, the dialogue manager can fall back to the current general question, and if this fallback fails, simply output a generic default response.

\subsection{Interface}
\begin{figure}
    \centering
    \includegraphics[width=8cm]{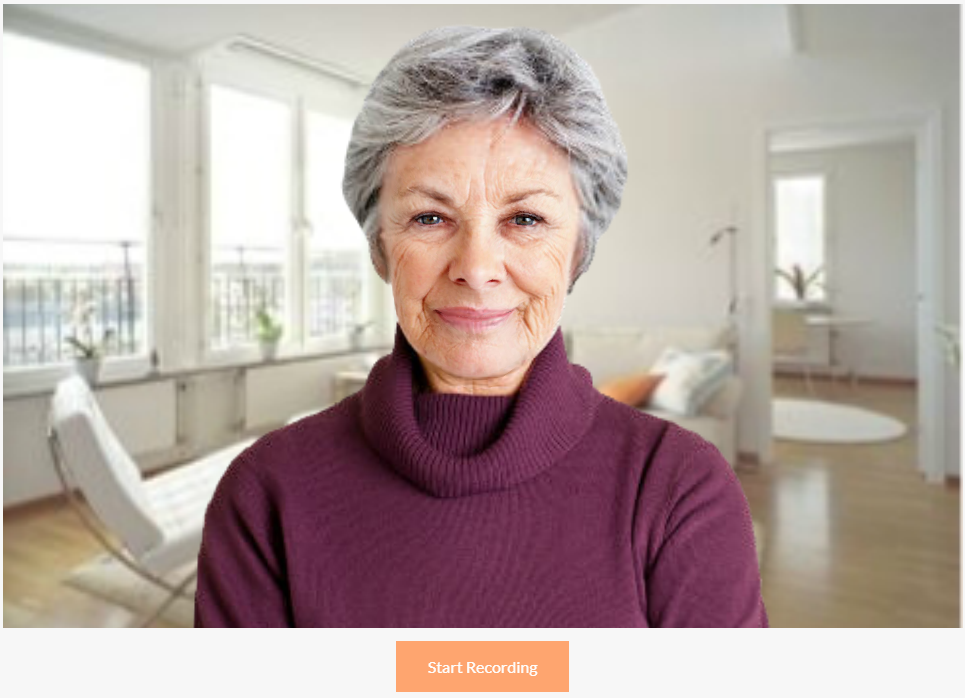}
    \caption{SOPHIE virtual agent}
    \label{fig:sophie}
\end{figure}

\begin{figure}[thb]
    \centering
    \includegraphics[width=14cm]{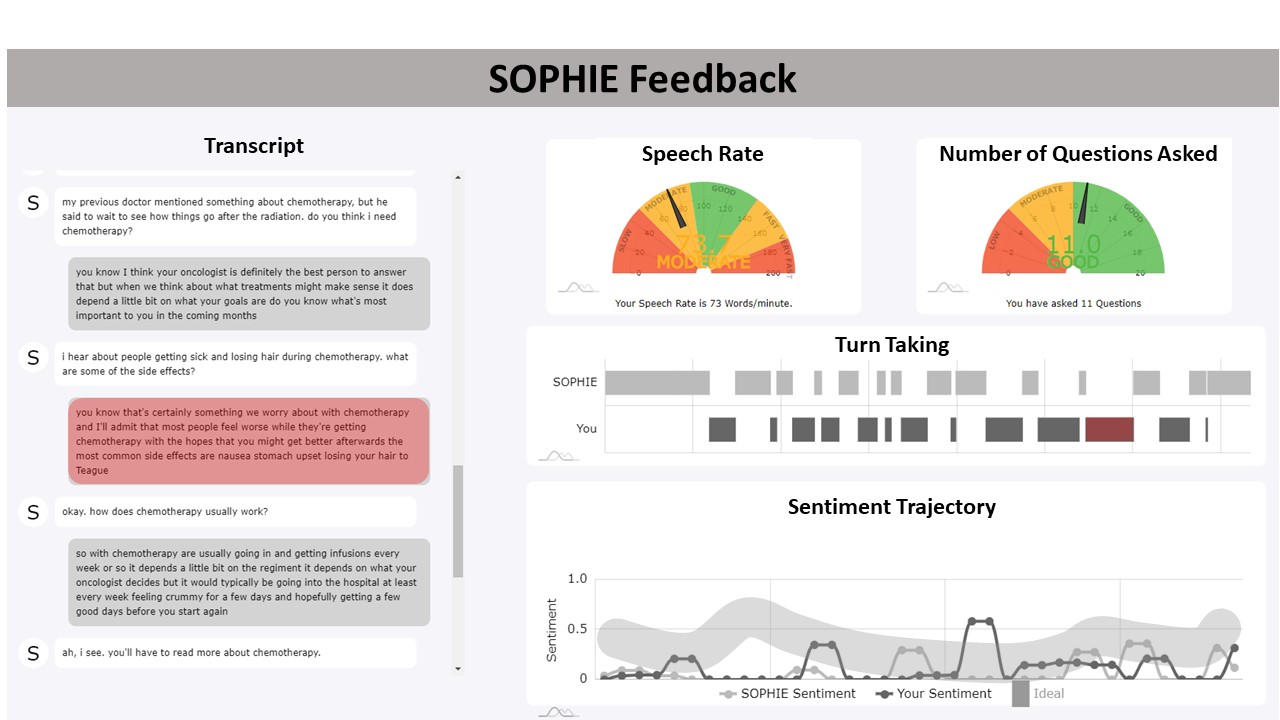}
    \caption{Feedback interface of SOPHIE. On the left side the conversation transcript is shown. On the right (from top to bottom) speech rate, number of questions are shown. Turn taking shows the turn length and at the bottom the sentiment trajectory of both physician and SOPHIE are shown with the ideal/suggested sentiment trajectory. }
    \label{fig:sophie_feedback}
\end{figure}

The SOPHIE system features a virtual agent (shown in fig. \ref{fig:sophie}). At the beginning, users start the conversation by pressing the "start recording" button. Users can then proceed to conversing with SOPHIE, and when the conversation is over the program takes the user to the feedback page. The feedback interface is shown in Fig. \ref{fig:sophie_feedback}. On the left side of the feedback interface we show the conversation transcript. The red marked speech is considered too long for the patients, i.e., it is classified as lecturing. On the right side of the feedback we show the speech rate of the user, the number of questions the user asked, turn taking, and the sentiment trajectory. Past literature has established that conversational speech rate is important in enabling patients to understand their prognosis. Also, asking questions of the patient is important for ensuring that the patient understands what is being said (\cite{back2005approaching}). The turn taking annotation shows the length of each turn by SOPHIE and the user. The example was chosen to illustrate the lecturing style of conversation; the detection of lecturing style was explained in section \ref{sec:lecturing}. The feedback shows the sentiment trajectory of both SOPHIE and the user. Additionally, the feedback shows a suggested sentiment trajectory for the user. The feedback page displays explanations of individual items when users hover their mouse on them. 

\subsection{Pilot Study}
To further assess acceptability and usability, we conducted a pilot study with eight practicing clinicians (fellows, residents, and nurse practitioners) from a University Medical Center. Our goal was to gather more information about their experience with SOPHIE, any limitations, and how we could improve the system. The study was performed with one participant at a time on the zoom communication platform. Each day, we asked the invited participant to have a conversation with SOPHIE and to look at the feedback. 

After conversing and receiving the feedback we interviewed the participants. The aim of the interview was to understand the accuracy and usefulness of the feedback, the appropriateness of the conversation, and suggestions for new features. We have performed a thematic analysis on the interview transcripts; our findings follow below. 

\subsubsection{Medical History}
All the participants mentioned that a brief medical history should be presented before starting the conversation with SOPHIE. One participant said,
\begin{quote}
    “I think some kind of medical record would be extremely helpful. I thought I don’t have any information to say to her.”
\end{quote}
The participants mentioned that in a regular standardized patient visit, they are given a medical record before they go into the room. They suggested the same scenario should be replicated for SOPHIE. In our program, SOPHIE starts the conversation by mentioning her increasing pain. The participants felt that this was abrupt and there should be a transition to this serious topic. They also mentioned that the way SOPHIE initiated presentation of her symptoms was unusual. In most cases, patients do not actively start the conversation. Rather, the physician looks at the patient's medical record and then starts asking about any new symptoms. In future we expect to modify the dialogues so that SOPHIE appears more passive and lets the users ask questions, though completely user-driven conversation remains beyond the state of the art. 

\subsubsection{Topics of Conversation}\label{sec:conversation_topics}
Participants (four out of eight) mentioned that SOPHIE jumped between topics and did not allow full coverage of a given topic. For example, SOPHIE begins talking about her pain medication, but the participants often asked questions about the current dosage and about other pain medication she had taken in the past. Since SOPHIE's limited dialogue repertoire falls short of covering those questions, she starts talking about her current medication (i.e., Lortab) and then about her test results. One participant said, 
\begin{quote}
    “The dialogue didn’t match with the questions I was asking. When she mentioned pain and I was trying to find more about the pain in order to help her with her question. But the answers that I gave her to her questions did not really fit and she just jumped to the next topic so I jumped with it but that was a little bit jarring to me.”
\end{quote}

Although SOPHIE changed the conversation topics abruptly, the questions she asked were found to be realistic. Five participants felt that SOPHIE was able to express her concerns and make them feel the seriousness of the situation. One participant added,

\begin{quote}
     “I think the topics were absolutely realistic. All the questions she asked were appropriate.”
\end{quote}

\subsubsection{Feedback on Speech Rate}
Participants (seven out of eight) mentioned that the speech rate feedback was easy to understand and very useful. One participant said, 
\begin{quote}
    "I know I tend to speak very fast, receiving feedback on my speech rate is going to be very useful."
\end{quote}
Another participant mentioned that in normal practice there is no way of measuring the speech-rate. However, with SOPHIE we could provide the information about how fast the physicians are speaking, which is useful.
\begin{quote}
    “I think the feedback (speech-rate) was useful. I never had someone measure my speech rate before. Sometimes I try to be cognizant of speaking a little bit slower with the patients but it was nice to actually get some feedback like you are doing okay.”
\end{quote}

One participant mentioned that it is important to speak more slowly when delivering bad news. She said,
\begin{quote}
    “For me it (speech-rate feedback) is useful, because I know that I have a tendency to speak really fast. So especially when I am delivering bad news I try to be super cognizant. ”
\end{quote}

However, the participants also noted that SOPHIE's speech rate was constant, making it difficult for them to adjust their speech rate depending on whether they are discussing serious issues or a casual topic. In the future, we plan to adjust SOPHIE's speech rate based on the seriousness of the topic being discussed.

\subsubsection{Number of Questions Asked}
Seven out of eight participants expressed that feedback on the number of questions asked was very useful. One participant said, 

\begin{quote}
    “It was helpful to get the information about how many questions you have asked, because I think a lot of the times we walk away from the conversation thinking that we really invited the patients into the talk, when maybe we didn't and we did a lot of lecturing. So I think that was a valuable feedback.”
\end{quote}
In addition to the number of questions asked, participants suggested that it should be highlighted what type of questions were asked, for example, how many history-taking questions were asked and how many emotional questions were asked. Though they expressed mixed feelings, participants (six out of eight) stated that this feedback would encourage them to ask more questions in the future. 

\subsubsection{Explanation of Sentiment}
The participants asked for more explanation on the sentiment trajectory. Seven participants mentioned that they did not understand the meaning of the sentiment values. They also said that the sentiment feedback is hard to interpret and they often confused it with empathy. Four participants wanted to see an example sentence of positive and negative sentiment. The participants also mentioned that changing or adjusting sentiment while engaged in the conversation may add to the cognitive load. They suggested that instead of asking the user to be positive at certain moments we should just stress the importance of dynamically adjusting sentiment. 

\subsubsection{Additional Feedback}
The participant also asked to add some additional feedback that they found useful in practice. Two participants said that there are few expressions of empathy in the dialogue and they should be highlighted in transcripts so that users could look back and understand how they responded to them. One participant suggested we should give feedback on the way users addressed concerns. 

One participant said that the turn-taking feedback is useful, however, it does not show the total amount of time a person was speaking. The participant said,  
\begin{quote}
    "I tend to speak a lot, but I don't want to make the patients feel that I am not listening. I want to know that I am giving a chance to ask questions."
\end{quote}
He suggested addition of a bar chart to the feedback page that indicates the total speech times for SOPHIE and the user. 

Three participants suggested giving feedback on nonverbal behaviors, such as eye contact. One of them said,
\begin{quote}
    “One of the things I think is important, and I have seen it in other clinicians, is eye contact. I think it's super important when we are giving bad news or having difficult conversations. I have colleagues who tend not to look at the patients”.
\end{quote}

\section{Discussion, Limitations, and Future Work}

We have described two novel contributions to communication research; empiric associations between automatically detected behaviors and patient prognostic understanding, and the development of SOPHIE, an automated system for teaching and evaluating patient-physician communication. In addition to the communication training program, the automatic detection of behaviors can be applied in pre-recorded standardized patients interaction to evaluate the communication skills. We acknowledge some limitations in the development of SOPHIE and the use of such a system as a basis for feedback. 

First, it should be noted that our finding of associations between trajectory styles and lecturing tendencies with prognosis understanding measures may not be causal. Our lecturing analysis was motivated from prior research that suggested that when a physician tends towards lecturing, it results in the patient not retaining as much of the information presented \cite{back2005approaching}. An alternative explanation could be that when physicians sense that patients do not understand, physicians are motivated to speak more, explain in greater detail, leading to a more lecturing-like structured conversation. Additionally, apparently passive patients may just lack understanding, which can result in poor engagement (i.e., patients may be too embarrassed or confused to ask for clarification), and this may result in conversations with a high LECT-UR score. 

In explaining the association of higher prognosis understanding with the dynamic sentiment trajectory style, we surmise that being dynamic keeps the patient more engaged, and that ending on a positive note keeps the patient less depressed and more likely to remember the information just presented. However, again, an alternative \textit{anticausal} explanation could be that patients' lack of prognosis understanding, and their physician's perception of this, motivates the physician to speak in a calmer, less dynamic way (e.g., sentiment trajectory styles B or C).

Additionally, the extent that the LECT-UR score correlates with human annotated instances of "ground truth" lecturing should be investigated. However, it should be noted that despite any difference between the LECT-UR lecturing-like structure measure and human-labelled ground truth instances of lecturing, our results establish that the LECT-UR score serves as a useful metric in its association with patient prognosis misunderstanding.

Some limitations exist with regards to the bigger picture of SOPHIE-like virtual agents. Past research suggests that while conversing with a virtual agent or AI driven conversational agent, humans tend to use shorter turns \cite{hill2015real}. This could be a limitation of using SOPHIE to train users to avoid lecturing, since users might use shorter turns regardless of feedback. Our LECT-UR scoring method utilizes a window of consecutive turns that also includes the virtual agent's turn. This allows the lecturing feedback to dynamically adopt to the conversation states and to the user's behavior. We think that this can help circumvent the limitation posed by using feedback trained on human-human conversation with a computerized dialogue system, though addressing this concern through a randomized study remains part of our planned future work.

The current dialogue manager itself also has some limitations, which we aim to address in the future. First, the output of the currently used automatic speech recognition (ASR) software\footnote{https://www.nuance.com/index.html} does not include punctuation. This limits the agent's ability to interpret the user correctly; for example, the pattern transduction mechanism would detect questions more reliably if they ended in an explicit question mark. Secondly, as discussed in Section \ref{sec:conversation_topics}, the dialogue manager tended to abruptly jump to the next topic in the main dialogue schema in cases where it failed to understand the user's input. This will be addressed by further expanding the interpretation patterns on the basis of the dialogues we observed in this study, as well as by allowing for more robust default strategies, such as staying on topic when it appears that the agent misinterpreted the user's input or when the user's input appears irrelevant to the agent's question.

While our study focused on high patient prognosis understanding as a positive goal, it should be acknowledged that patients sometimes don't want to know specifically how much time they have left \cite{rodenbach2017promoting}. In designing a communication training program we should incorporate options as to how much information the physician should deliver. Another limitation of this work may be that our findings are limited to patient-physician relationships involving diseases and conditions as serious and sensitive as advanced cancer care and end-of-life communication.

Regarding the analysis of sentiment trajectories, we found that three clusters represent the data best according to an information-theory-based metric. However, it is important to verify that the data is not better represented by a single cluster (k=1) (i.e. no clustering), rather than multiple clusters (k >= 2). Because the Silhouette score method can only be applied with k >= 2, we use the Bayesian Information Criterion (BIC) \cite{schwarz1978estimating} as an additional method of finding the optimal k which also allows us to determine that separate clusters do not exist. More specifically, the BIC method is applicable when the clusters are represented probabilistically (i.e. with a probability density function) which is not provided by the k-means algorithm. We thus used a related clustering technique, the Gaussian Mixture Model (GMM) to determine whether the data is better represented by a single cluster. While it is possible to use a Gaussian Mixture Model as our primary clustering method instead of k-means, there are multiple reasons why k-means is more appropriate. First, the distribution of sentiment values was skewed, whereas skewed data cannot be represented with a Gaussian distribution. Second, the sentiment fall into the fixed interval [0, 1], unlike a Gaussian which spans $[-\infty , \infty]$. 
In addition to considering the number of clusters we have experimented with a range of values for the number of segments. A large segment is not suitable for trajectory analysis since it may contain the bulk of the conversation, and a small segment size is also not suitable since it may not contain representative turns from both physicians and patients. Thus we experimented with five, eight, ten, and fifteen as our number of segments. In this paper, we have shown results for the choice of eight segments, omitting the others as they produced similar results.  

Despite these limitations, SOPHIE in its current form served as a starting point for developing the communication skills program for physicians. The pilot study allowed us to identify the areas where we should make further modifications. In future versions of SOPHIE we also plan to incorporate the suggestions made by the clinicians.  

\section{Conclusion}
In summary, in this paper, we provide early results of our multi-stage research examining patient-physician conversations, identification of effective traits (not lecturing, asking questions, delivering news on a positive note), development of an automated way of evaluating these traits, and the design of a real-time online standardized patient-physician communication training system where an avatar plays the role of a standardized patient. We structured our exploration in the context of conversations between final stage cancer patients (i.e., terminal patients) and oncologists. 

We believe that successful SOPHIE-like systems could have broader global impacts. Two thirds of the cancer deaths happens in the low and mid income countries such as countries in Latin America and sub-Saharan Africa \cite{stoltenberg2020central,van2020increased}. However, most of the serious ill patients don’t have access to the quality palliative care (PC) due to the inadequate PC training programs. Current medical training in the countries of these regions focuses on treating diseases. The comfort care in chronic life threatening diseases such as cancer, is still in its infancy. In Africa, Kenya, Uganda and Botswana have initiated post-graduate training programs for palliative care \cite{malloy2017providing,kamonyo2018palliative}. Only South Africa has a well-established post-graduate and research program on palliative care\cite{gwyther2007palliative}. We are hopeful that online programs such as SOPHIE can provide a basis for helping these communities develop training programs for PC physicians.  

\bibliographystyle{ACM-Reference-Format}
\bibliography{bib}

\end{document}